# Spin pumping as a generic probe for linear spin fluctuations: demonstration with ferromagnetic and antiferromagnetic orders, metallic and insulating electrical states

Olga Gladii,[1,*] Lamprini Frangou,[1,*] Guillaume Forestier,[1] Rafael Lopes Seeger,[1,2] Stéphane Auffret,[1] Miguel Rubio-Roy,[1] Raphael Weil,[3] Alexandra Mougin,[3] Christelle Gomez,[4] Walaa Jahjah,[5] Jean-Philippe Jay,[5] David Dekadjevi,[5] David Spenato,[5] Serge Gambarelli,[6] and Vincent Baltz[1,**]

[1] Univ. Grenoble Alpes, CNRS, CEA, Grenoble INP, INAC-Spintec, F-38000 Grenoble, France
[2] Departamento de Física, UFSM, Santa Maria, 97105-900 Rio Grande do Sul, Brazil
[3] LPS, Univ. Paris-Sud / Univ. Paris-Saclay / CNRS, F-91405 Orsay, France
[4] Grenoble INP / CIME Nanotech, F-38000 Grenoble, France
[5] OPTIMAG Univ. Bretagne Occidentale, F-29238 Brest, France
[6] SYMMES, Univ. Grenoble Alpes / CNRS / INAC-CEA, F-38000 Grenoble, France
[*] equal contribution
[**] vincent.baltz@cea.fr

**Abstract**

We investigated spin injection by spin pumping from a spin-injector(NiFe) into a spin-sink to detect spin fluctuations in the spin-sink. By scanning the ordering-temperature of several magnetic transitions, we found that enhanced spin pumping due to spin fluctuations applies with several ordering states: ferromagnetic(Tb) and antiferromagnetic(NiO, NiFeOx, BiFeO$_3$, exchange-biased and unbiased IrMn). Results also represent systematic experimental investigation supporting that the effect is independent of the metallic and insulating nature of the spin-sink, and is observed whether the spin current probe involves electronic or magnonic transport, facilitating advances in material characterization and engineering for spintronic applications.

Spintronics relies on the spin-dependent transport properties of matter. In this field, spin currents [1] are key to facilitating characterization and engineering of new materials. The spin pumping effect [2] has attracted considerable attention due to its versatile capacity to unravel a variety of spin-dependent transport phenomena, such as interface spin-filtering [3], spin diffusion length [4], spin-charge interconversion [5–7], and spin fluctuations [8–10].

The spin pumping mechanism involves the injection of a spin current from an out-of-equilibrium ferromagnetic spin-injector into an adjacent layer known as the spin-sink, resulting in the loss of spin angular momentum. This loss enhances the rate at which the ferromagnet relaxes toward equilibrium [2]. In practice, the out-of-equilibrium magnetization dynamics of the spin-injector is most often driven by ferromagnetic resonance. Measurement of Gilbert damping ($\alpha$) [11,12] is used to characterize the loss of spin angular momentum in the spin-sink. Gilbert damping can be described as the sum of local damping ($\alpha^0$) due to intraband and interband scattering [13] and non-local damping ($\alpha^p$) associated with the loss of angular momentum due to spin pumping. Since spin injection is related to a quality called spin mixing conductance ($g^{\uparrow\downarrow}$), as framed in a theory involving adiabatic charge pumping [2], the non-local damping is also connected to this quality ($\alpha^p \propto g^{\uparrow\downarrow}$). This relation makes spin pumping an efficient method to probe spin fluctuations as they open new conduction channels across the interface, resulting in an enhanced $g^{\uparrow\downarrow}$. In other words, as initially presented theoretically in a linear-response formalism by Ohnuma et al. [8] describing spin pumping near thermal equilibrium, spin mixing conductance is linked to the dynamical transverse spin susceptibility of the spin-sink, $\chi_k^R$, through $g^{\uparrow\downarrow}(T) \propto \sum_k \frac{1}{\Omega_{rf}} \text{Im} \chi_k^R(\Omega_{rf}, T)$, where $k$ is the wave vector, T is the temperature, and $\Omega_{rf}$ is the angular frequency of the ferromagnetic spin-injector at resonance. Consequently, the non-local damping is connected to the dynamical transverse spin susceptibility of the spin-sink. Since this spin susceptibility is enhanced around most ordering transitions, spin pumping should generically result in the temperature-dependence of $\alpha^p$ reaching a maximum for all magnetic and electric spin-sinks.

Although the initial description of spin pumping near thermal equilibrium was formulated for a ferromagnetic spin-sink [8], there is currently no clear experimental demonstration in this particular case. The reason for this is because experimental application of the method proved to be more useful for antiferromagnetic spin-sinks [9,10,14] due to the absence until then of a benchtop technique to access paramagnetic-to-antiferromagnetic transitions in thin films, as for example pointed out early in Ref. [15]. While a paramagnetic to ferromagnetic phase transition can be recorded from simple magnetometry experiments, by measuring the temperature-dependence of magnetization, the magnetic phase transition of an antiferromagnet lacks net magnetization and is therefore not accessible in this way. Alternative techniques using local probes such as neutrons [16] are also unsuitable for films of antiferromagnetic material a few nanometers thick, since their signal-to-noise ratio is limited by the small volume. The first experimental demonstrations of spin pumping as a spin fluctuation and susceptibility probe were presented simultaneously by Qiu et al. [10] for the case of coupled and uncoupled (through a Cu spacer) CoO and NiO antiferromagnetic insulators and by Frangou et al. [9] for the case of an uncoupled (through a Cu spacer) IrMn antiferromagnetic metal in a fully metallic stack. Published studies can be split into three different cases [14]: first, in ferromagnet/non-magnet/antiferromagnet metallic trilayers, spin transport is purely electronic through the non-magnetic metal, i.e., spins are carried by conduction electrons; second, in exchange-biased ferromagnet/antiferromagnetic-insulator bilayers, transport is purely magnonic, i.e., due to excitation of localized-magnetic-moments; and third, in exchange-biased ferromagnet/antiferromagnet metallic bilayers, both electronic and magnonic transport regimes may coexist since transport by conduction electrons is permitted while magnons produced simultaneously by the oscillating ferromagnet feed directly into the antiferromagnet due to exchange bias interactions. Whether the nature of the probe, i.e., the magnonic vs. electronic nature of the spin current injected and absorbed in the spin-sink, influences the efficiency of damping enhancement near the magnetic phase transition is still a subject of debate. Beyond spin fluctuations in ferromagnets and antiferromagnets, the initial formalism of spin pumping near a phase transition was recently theoretically extended to the case of normal to superconducting transitions by assessing the dynamic spin susceptibility of a superconductor [17], reinforcing interest in investigating various types of magnetic ordering.

Here, we examined temperature-dependent ferromagnetic relaxation in thin NiFe films and how it was affected by spin fluctuations in adjacent spin-sinks with a range of ordered and electrical states: ferromagnets (Tb); antiferromagnets (NiO, NiFeOx, $BiFeO_3$, exchange-biased IrMn, and unbiased IrMn); metals (Tb, IrMn); and insulators (NiO, NiFeOx, $BiFeO_3$). Our results represent systematic experimental support that the technique is generic and functions regardless of whether the probe involves spin-wave-like or electronic-like transport.

The full stacks used in this study were (from substrate to surface): NiFe(8)/Cu(3)/Tb(3)/Al(5), NiFe(8)/Cu(3)/IrMn(0.6)/Al(2), NiFe(8)/IrMn(0.6)/Al(2), Ta(3)/NiO(1.5)/NiFe(7)/Cu(3), NiFe(8)/NiFeOx(1.5), and Ta(15)/$BiFeO_3$(3)/NiFe(8)/Ta(3) multilayers. All thicknesses are given in nanometers. Stacks were deposited on thermally oxidized silicon substrates [Si/$SiO_2$(500)] at room temperature. The Tb-based sample was deposited by molecular beam epitaxy, all others were produced by dc-magnetron sputtering. The NiFe layer was deposited from a $Ni_{81}Fe_{19}$ (at. %) permalloy target. The IrMn layers were deposited from an $Ir_{20}Mn_{80}$ (at. %) target. An Al cap was added when necessary to block oxidization by air, it formed a protective passivating AlOx film. Uncapped NiFe formed a passivating 1.6-nm thick NiFeOx layer [18]. A 3-nm thick Cu layer was used in some samples to break the direct magnetic interaction between the spin-injector and the spin-sink. Six stacks consisting of similar multilayers without Tb, IrMn, NiO, NiFeOx, and BiFeO$_3$ spin-sinks, respectively, were also systematically deposited. They were used as references to further isolate the spin-sink contribution to Gilbert damping, as detailed below. The spin-sink thicknesses were chosen to give a magnetic phase transition within the temperature range accessible in our experimental setup [19].

Spin pumping experiments (Fig. 1(a)) and the corresponding ferromagnetic resonance spectra (Fig. 1(b)) were recorded at temperatures (T) ranging between 5 and 300 K, using a continuous-wave electron paramagnetic resonance spectrometer operating at 9.6 GHz and fitted with a cavity. For each temperature tested, the total Gilbert damping ($\alpha$) was determined by fitting the NiFe differential resonance spectrum to a Lorenzian derivative (Fig. 1(b)). The value of $\alpha$ is given by $\alpha(T) = \left[\Delta H_{pp}(T) - \Delta H_0(T)\right]\sqrt{3}|\gamma|/(2\omega)$, where $\Delta H_{pp}$ is the peak-to-peak linewidth for the spectrum, $\Delta H_0$ is the inhomogeneous broadening due to spatial variations in the magnetic properties, $\gamma$ is the gyromagnetic ratio, and $\omega$ is the angular frequency [12]. For every system, the temperature-dependence of the NiFe Gilbert damping in the absence of influence of the spin-sink, $\alpha^{ref}(T)$, was deduced from measurements performed with the reference sample (Fig. 1(c)). The temperature-dependence of $\alpha^p$ was calculated using interpolation functions, by subtracting $\alpha^{ref}(T)$ from the value of $\alpha(T) = \alpha^{ref}(T) + \alpha^p(T)$ (Fig. 1(c)). Note that $\Delta H_0$ was found negligible at room temperature from measurements of standard $\Delta H_{pp}$ vs. $\omega/(2\pi)$ plots (Fig. 1(d)) using a separate broadband coplanar waveguide. It was also found to be a temperature-invariant parameter with a mean value of about 2 Oe, i.e., tenfold smaller than $\Delta H_{pp}$ [9]. Also note that the small value of $\Delta H_0$ that we measure is likely an upper bond since it also contains the contribution of the spatial distribution of the microwave magnetic fields. We considered that it can be eliminated from the calculation of $\alpha^p$.

We will first discuss the results obtained for a ferromagnetic spin-sink: a 3-nm-thick Tb layer [20] in a NiFe(8)/Cu(3)/Tb(3)/Al(5) multilayer. Figure 2(a) shows an enhanced spin angular momentum relaxation rate ($\alpha^p$) near 40 K (Left axis). The link between enhancement of $\alpha^p$ and spin fluctuations in Tb due to the paramagnetic to ferromagnetic phase transition undoubtedly correlates with the onset of saturation of the Tb layer's magnetization, as measured by magnetometry (Right axis and inset). The additional smooth temperature-dependence of the saturation magnetization is related to the NiFe(8) spin-injector. Data-fitting using the Bloch equation: $M_{S,NiFe}(T) = M_{S,NiFe}(0)\left(1-\beta T^{3/2}\right)$ returned $M_{S,NiFe}(0) = 785$ emu.cm$^{-3}$, and $\beta = 1.7$ x 10$^{-5}$ K$^{-3/2}$ (see line in Fig. 2(a)). These results are in satisfactory agreement with expectations. Although the initial description of spin pumping near thermal equilibrium was formulated for a ferromagnetic spin-sink, it should be remembered that clear experimental demonstration was missing until now [8,21,22]. We also note that the reduction of the Curie temperature from 200 K for bulk Tb [23] down to 40 K for a 3-nm thick layer can be ascribed to known finite-size effects [19,24]; finite-size scaling of ordering temperatures will be discussed below. The values of $\alpha^p$ at room temperature ($\alpha^p_{300K}$) and at the ordering transition temperature ($\alpha^p_{T=Tc}$) are listed in table 1. Paramagnetic Tb(3) is known to be a poor spin-sink, which explains why the value of $\alpha^p_{300K}$ is practically equal to zero [25]. Values listed in table 1 for a number of stacks will be compared and discussed below. However, first we feel it is important to briefly comment on the temperature-dependence of the resonance field, $H_{res}(T)$ (Fig. 2(b)). The high-temperature behavior can be satisfactorily described using the usual Kittel formula [26]:

$$\omega = |\gamma|\sqrt{\left(H_{res}(T)+H_K\right)\left(H_{res}(T)+H_K+4\pi M_s^{eff}\right)}, \qquad \text{where}$$

$M_s^{eff}(T) = M_{S,NiFe}(T) - 2K_{S,NiFe}/\left(4\pi M_{S,NiFe}(T)t_{NiFe}\right)$ is the effective magnetization, $t_{NiFe}$ is the thickness of the NiFe layer, $K_S$ is the surface anisotropy, and $H_K$ is the effective field resulting from volume anisotropy. This final parameter is known to be of the order of a few Oe, and can be neglected. Data-fitting (line in Fig. 2(b)) returned $M_{S,NiFe}(0) = 800$ emu.cm$^{-3}$, $\beta = 1.3 \times 10^{-5}$ K$^{-3/2}$, and $K_S = 0.5$ erg.cm$^{-2}$, which are in satisfactory agreement with the values obtained from the Bloch fit in Fig. 2(a). Interestingly, the shape of the low-temperature region is not satisfactorily described by the bare formula. Dipolar coupling between the NiFe and Tb layers through the Cu spacer could explain why the NiFe resonance field deviates from the usual Kittel equation. In fact, since the ferromagnetic Tb layer is not saturated around $H_{res}$ ~1 kOe (see inset in Fig. 2(a)), it contains domains and consequently results in a stray field ($H_{Tb}$) with a non-zero in-plane projection which is sensed by the NiFe layer. This field is related to the magnetization of the Tb layer ($H_{Tb} \propto 4\pi M_{1kOe,Tb}$) and it contributes as an effective field to the Kittel equation. Its influence becomes more marked when $M_{1kOe,Tb}$ increases as the temperature decreases, resulting in the gradual reduction of H$_{res}$ observed. The inset in Fig. 2(b) demonstrates that $M_{1kOe,Tb}$, deduced from hysteresis loop measurements (inset in Fig. 2(a)), and the deviation from $H_{res}$ are proportional.

Figure 3 shows $\alpha^p$ plotted against temperature for antiferromagnetic spin-sinks. We first compared two cases for the same metallic antiferromagnet without and with a Cu spacer. In the NiFe(8)/IrMn(0.6)/Al(2) multilayer, the IrMn spin-sink is directly fed by spin-waves through direct magnetic coupling with the NiFe spin-injector, and by conduction electrons as well. In contrast, in the NiFe(8)/Cu(3)/IrMn(0.6)/Al(2) multilayer [9], the spin current flows via conduction electrons through Cu and the Cu breaks the direct magnetic interaction between the IrMn and NiFe layers. The potential magnonic contribution to the spin current in the IrMn layer is therefore the result of electron-magnon conversion mechanisms and is probably less efficient than direct feeding. The dynamical transverse spin susceptibility of antiferromagnets is known to be enhanced when spins fluctuate [9], similar to ferromagnets. As a consequence, $\alpha^p$ reaches a maximum near the paramagnetic-to-antiferromagnetic phase transition of the IrMn layer, which we measured at T = 55 K (Fig. 3(a)). A similar maximum is reached in both the no Cu spacer and Cu spacer cases. Since the relative amplitude of the maxima, $\alpha^p_{55K}/\alpha^p_{300K}$, is the same for both cases (showing a 15-fold increase, table 1) we conclude that the phenomenon of spin pumping enhancement near a phase transition is independent of the Cu spacer, in agreement with Ref. [10]. It will be further confirmed later in the text that it is also independent of the type of transport - spin-waves-like or conduction-electron-like. The relative enhancement of spin mixing conductance due to spin fluctuations at the phase transition ($g^{\uparrow\downarrow}_{T=Tc}/g^{\uparrow\downarrow}_{300K}$) is listed in table 1 for several cases. This ratio can be approximated from $\alpha^p_{T=Tc}/\alpha^p_{300K} = g^{\uparrow\downarrow}_{T=Tc}/g^{\uparrow\downarrow}_{300K}$ [2]. We note that $\alpha^p_{300K}$ is 10 times larger for the sample without Cu spacer (2 vs. 0.2, see table 1). This effect is probably the result of direct exchange bias coupling, meaning that although the transport regime does not influence the relative enhancement due to spin fluctuations, it does influence the initial value at room temperature [14]. The data in table 1 also show that larger $\alpha^p_{300K}$ are recorded with magnonic transport in several kinds of materials. This effect may be the consequence of deeper penetration of the spin current carried by magnons compared to that flowing via conduction electrons [27,28]. In Fig. 3(b) it should be noted that the resonance field is temperature-dependent for the exchange-biased sample but not for the non-exchange-biased one. This behavior is known to result from rotatable anisotropy [28,29], i.e., from the presence of uncompensated spins in the antiferromagnet which have a longer relaxation time than the characteristic time for ferromagnetic resonance (~10 ns). These spins are dragged by the ferromagnet in a quasistatic experiment (~10 min) but stay still in a dynamic experiment, resulting in additional anisotropy for the ferromagnet. $H_{rot}(T)$ (inset in Fig. 3(b)) was calculated using the Kittel equation, replacing $H_K$ by $H_K + H_{E,st}(T) + H_{rot}(T)$, where $H_{E,st}(T)$ is the temperature-dependence of the hysteresis loop shift, recorded separately using a quasistatic magnetometer.

We note that several experiments now demonstrate that that two-magnon scattering can be ruled out: the position of the bump is frequency-independent [10], it superposes for the in-plane and out-of-plane configurations [28], and it corroborates with the ordering transition temperature determined separately by X-ray magnetic linear dichroism [10] and calorimetry [9]. For the NiFe/Cu/IrMn sample, despite the occurrence of a damping maximum, the temperature-independent behavior of $H_{res}$ is another good indication that the process does not involve paramagnetic relaxation.

We next investigated insulating antiferromagnets, a 1.5-nm thick NiO, a 1.6-nm thick NiFeOx and a 3-nm thick $BiFeO_3$ layer (Fig. 4(a)), in which spin current is carried by spin-waves. We found enhanced spin pumping near the paramagnetic-to-antiferromagnetic transition, at T = 85, 65, and 20 K, respectively. This result further underlines that spin current carried by spin-waves or by conduction electrons both efficiently reveal enhanced spin pumping due to spin fluctuations, in addition to experimentally supporting the universality of the phenomenon. It is also noteworthy from the results presented in table 1 that, overall, the spin mixing conductance ratio stays within the same order of magnitude, regardless of the nature of the ordering transition. With regard to the position of the spin pumping maximum, it should be remembered that we purposely tuned the transition temperature to the temperature range accessible in our experimental setup by choosing appropriate spin-sink thicknesses. Indeed, the thickness-dependence of the ordering temperature is well described by theoretical models [19,24]. A typical example is plotted in Fig. 4(b) for the Néel temperature of NiO, comparing data from the present study to data from the literature [10,30–35]. Taking all data points into account, the fit using $(T_{N,bulk} - T_N(t_{NiO}))/T_N(t_{NiO}) = (t_{NiO}/\xi_0)^{-\lambda_{eff}}$ [19,24] gave an extrapolated correlation length at T=0 K $\xi_0 = 1.7$ +- 0.1 nm (about 4 monolayers). For this fit, we took $T_{N,bulk} = 520$ K [14] and the effective shift exponent for a three-dimensional Heisenberg antiferromagnet $\lambda_{eff} = 3$. This exponent corresponds to a critical exponent $1/\nu = 1.4$, as predicted by field theory from the three-dimensional $O(n)$ vector model after finite-size corrections [24]. It should be noted that data points were for non-identical stacks and recorded using different techniques. These differences may explain the level of discrepancy observed. Nevertheless, the overall thickness-dependent behavior was satisfactory. In addition, it is clear from Fig. 4(b) that x-ray and calorimetry techniques are suitable for measuring $T_N$ for thick layers, whereas spin pumping and spin Hall magnetoresistance have made it possible to explore more systematically the thin-layer regime (here sub-2 nm). Finally, if we return to Fig. 3(a), we can see that the position of the peak is not altered by exchange bias coupling, agreeing with the idea that the peak is an indicator of the ordering transition temperature.

In conclusion, the main contribution of this paper is that it represents systematic experimental investigation supporting the generic character of the phenomenon of enhanced spin pumping resulting from spin fluctuations in a spin-sink layer. The phenomenon was found to apply with all kinds of ordering and electrical states, regardless of the electronic or magnonic nature of the spin current probe. These results will facilitate progress in characterization and engineering of new materials.

**Acknowledgments**

We acknowledge financial support from ANR [Grant Number ANR-15-CE24-0015-01] and KAUST [Grant Number OSR-2015-CRG4-2626]. We also thank M. Gallagher-Gambarelli for critical reading of the manuscript.

# References

[1] S. Maekawa, S. O. Valenzuela, E. Saitoh, and T. Kimura (eds.), Ser. Semicond. Sci. Technol. Oxford Univ. Press. Oxford (2012).

[2] K. Ando, Semicond. Sci. Technol. **29**, 043002 (2014).

[3] Y. Tserkovnyak, A. Brataas, G. E. W. Bauer, and B. I. Halperin, Rev. Mod. Phys. **77**, 1375 (2005).

[4] J. Bass and W. P. Pratt, J. Phys. Condens. Matter **19**, 183201 (2007).

[5] A. Hoffmann, IEEE Trans. Magn. **49**, 5172 (2013).

[6] J. Sinova, S. O. Valenzuela, J. Wunderlich, C. H. Bach, and T. Jungwirth, Rev. Mod. Phys. **87**, 1213 (2015).

[7] E. Saitoh, M. Ueda, H. Miyajima, and G. Tatara, Appl. Phys. Lett. **88**, 182509 (2006).

[8] Y. Ohnuma, H. Adachi, E. Saitoh, and S. Maekawa, Phys. Rev. B **89**, 174417 (2014).

[9] L. Frangou, S. Oyarzun, S. Auffret, L. Vila, S. Gambarelli, and V. Baltz, Phys. Rev. Lett. **116**, 077203 (2016).

[10] Z. Qiu, J. Li, D. Hou, E. Arenholz, A. T. NDiaye, A. Tan, K.-I. Uchida, K. Sato, Y. Tserkovnyak, Z. Q. Qiu, and E. Saitoh, Nat. Commun. **7**, 12670 (2016).

[11] L. Landau and E. Lifshitz, Phys. Z. Sowjetunion **8**, 153 (1935).

[12] T. L. Gilbert, IEEE Trans. Magn. **40**, 3443 (2004).

[13] K. Gilmore, Y. U. Idzerda, and M. D. Stiles, Phys. Rev. Lett. **99**, 27204 (2007).

[14] V. Baltz, A. Manchon, M. Tsoi, T. Moriyama, T. Ono, and Y. Tserkovnyak, Rev. Mod. Phys. **90**, 015005 (2018).

[15] P. Merodio, A. Ghosh, C. Lemonias, E. Gautier, U. Ebels, M. Chshiev, H. Béa, V. Baltz, and W. E. Bailey, Appl. Phys. Lett. **104**, 032406 (2014).

[16] T. Yamaoka, M. Mekata, and H. Takaki, J. Phys. Soc. Japan **31**, 301 (1971).

[17] M. Inoue, M. Ichioka, and H. Adachi, Phys. Rev. B **96**, 024414 (2017).

[18] L. Frangou, G. Forestier, S. Auffret, S. Gambarelli, and V. Baltz, Phys. Rev. B **95**, 054416 (2017).

[19] R. Zhang and R. F. Willis, Phys. Rev. Lett. **86**, 2665 (2001).

[20] A. Mougin, C. Dufour, K. Dumesnil, P. Mangin, and G. Marchal, J. Magn. Magn. Mater. **165**, 168 (1997).

[21] B. Khodadadi, J. B. Mohammadi, C. Mewes, T. Mewes, M. Manno, C. Leighton, and C. W. Miller, Phys. Rev. B **96**, 054436 (2017).

[22] Y. Ou, D. C. Ralph, and R. A. Buhrman, Phys. Rev. Lett. **120**, 097203 (2018).

[23] A. Michels, J. P. Bick, R. Birringer, A. Ferdinand, J. Baller, R. Sanctuary, S. Philippi, D. Lott, S. Balog, E. Rotenberg, G. Kaindl, and K. M. Döbrich, Phys. Rev. B **83**, 224415 (2011).

[24] M. Henkel, S. Andrieu, P. Bauer, and M. Piecuch, Phys. Rev. Lett. **80**, 4783 (1998).

[25] J. Yue, S. Jiang, D. Zhang, H. Yuan, Y. Wang, L. Lin, Y. Zhai, J. Du, H. Zhai, J. Yue, S. Jiang, D. Zhang, H. Yuan, and Y. Wang, AIP Adv. **6**, 056120 (2016).

[26] C. Kittel, Phys. Rev. **73**, 155 (1948).

[27] H. Saglam, W. Zhang, M. B. Jungfleisch, J. Sklenar, J. E. Pearson, J. B. Ketterson, and A. Hoffmann, Phys. Rev. B **94**, 140412(R) (2016).

[28] O. Gladii, L. Frangou, G. Forestier, R. L. Seeger, S. Auffret, I. Joumard, M. Rubio-Roy, S. Gambarelli, and V. Baltz, Phys. Rev. B **98**, 094422 (2018).

[29] J. McCord, R. Mattheis, and D. Elefant, Phys. Rev. B **70**, 094420 (2004).

[30] W. Lin, K. Chen, S. Zhang, and C. L. Chien, Phys. Rev. Lett. **116**, 186601 (2016).

[31] E. Abarra, K. Takano, F. Hellman, and A. Berkowitz, Phys. Rev. Lett. **77**, 3451 (1996).

[32] C. Boeglin, O. Ersen, M. Pilard, V. Speisser, and F. Kronast, Phys. Rev. B **80**, 035409 (2009).

[33] D. Hou, Z. Qiu, J. Barker, K. Sato, K. Yamamoto, S. Velez, J. M. Gomez-Perez, L. E. Hueso, F. Casanova, and E. Saitoh, Phys. Rev. Lett. **118**, 147202 (2017).

[34] D. Alders, L. Tjeng, F. Voogt, T. Hibma, G. Sawatzky, C. Chen, J. Vogel, M. Sacchi, and S. Iacobucci, Phys. Rev. B **57**, 11623 (1998).

[35] W. Lin and C. L. Chien, Phys. Rev. Lett. **118**, 067202 (2017).

**Figure captions**

Fig. 1. (a) The spin pumping experiment. (b) Representative series of differential absorption spectra ($d\chi''/dH$ vs. H, where H is the external dc bias field applied in the sample plane) measured at different temperatures (T). The lines were fit to the data using a Lorenzian derivative. (c) Representative temperature-dependence of the peak-to-peak linewidth ($\Delta H_{pp}$) for a $BiFeO_3$/NiFe bilayer and its NiFe reference. (d) Corresponding frequency-dependence of $\Delta H_{pp}$ measured at 300K.

Fig. 2. (a) Temperature-dependence of the extrinsic spin pumping contribution ($\alpha^p$) to the NiFe layer's total Gilbert damping due to a Tb(3) metallic ferromagnetic spin-sink (Left) and temperature-dependence of the sample's saturation magnetization ($M_s$) (Right), for a NiFe/Cu/Tb trilayer. The line corresponds to a fit to the high-temperature data, considering only the NiFe contribution and based on a Bloch equation. Inset: corresponding typical in-plane hysteresis loops measured at several temperatures. (b) Temperature-dependence of the NiFe layer resonance field ($H_{res}$). The line corresponds to a fit to the high-temperature data using the Kittel equation. Inset: relationship between the deviation from $H_{res}$ (ascribed to the stray field, $H_{Tb}$, created by the unsaturated Tb layer) and the Tb layer's remanent magnetization ($M_{r,Tb}$). The line gives a linear fit and is constrained to pass through (0,0).

Fig. 3. (a) Temperature-dependence of $\alpha^p$ due to IrMn(0.6) metallic antiferromagnetic spin-sinks for a NiFe/IrMn bilayer (electronic and magnonic transport throughout) compared to a NiFe/Cu/IrMn trilayer (purely electronic transport through Cu). (b) Corresponding temperature-dependence of $H_{res}$. Inset: temperature-dependence of the static hysteresis loop

shift ($H_{E,st}$) and of the rotatable anisotropy contribution that contributes to the dynamic regime ($H_{rot}$).

Fig. 4. (a) Temperature-dependence of $\alpha^p$ due to NiO(1.5), NiFeOx(1.6), and BiFeO$_3$(3) insulating antiferromagnetic spin-sinks. (b) Thickness-dependence of the Néel ordering temperature for NiO. $C_p$ refers to calorimetry; XMCD and XMLD refer to x-ray magnetic circular and linear dichroism, respectively; SP refers to spin pumping; and SMR refers to spin-Hall magnetoresistance. Lines were fitted using the finite-size model described in the text.

Table 1. NiFe Gilbert damping at room temperature for the reference sample, $\alpha^{ref}_{300K}$ (without the spin-sink), spin-sink contribution to Gilbert damping, at room temperature, $\alpha^p_{300K}$, and at the phase transition, $\alpha^p_{T=Tc}$, and corresponding spin mixing conductance ratio, $\frac{g^{\uparrow\downarrow}_{T=Tc}}{g^{\uparrow\downarrow}_{300K}}$. Data are listed for spin-sink layers of several magnetic and electronic kinds. Note: the samples and their respective references were deposited using three different machines (see gray separations).

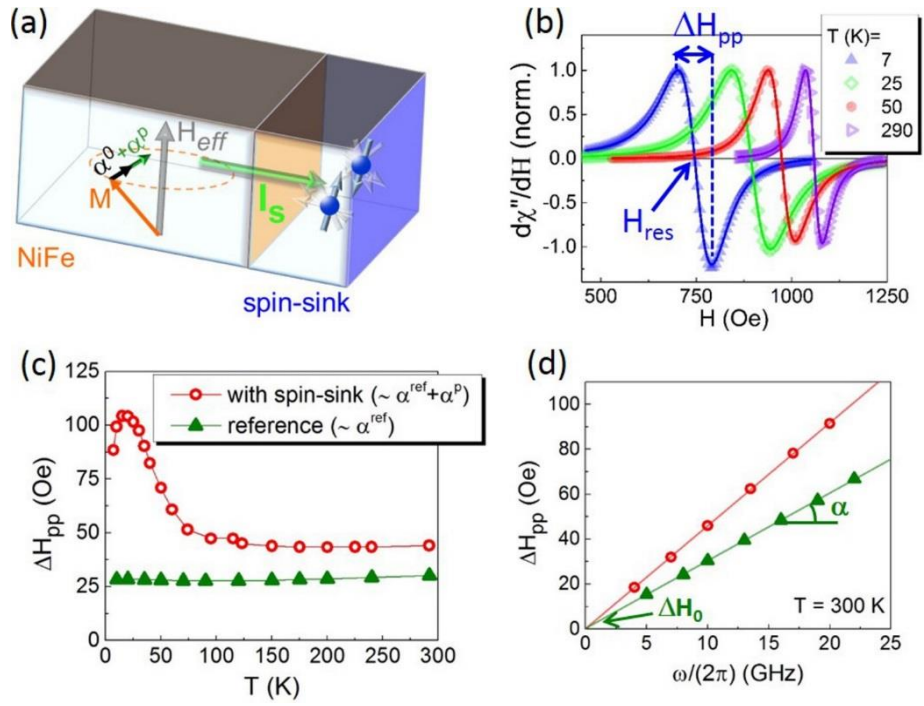

O. Gladii et al    Fig. 1

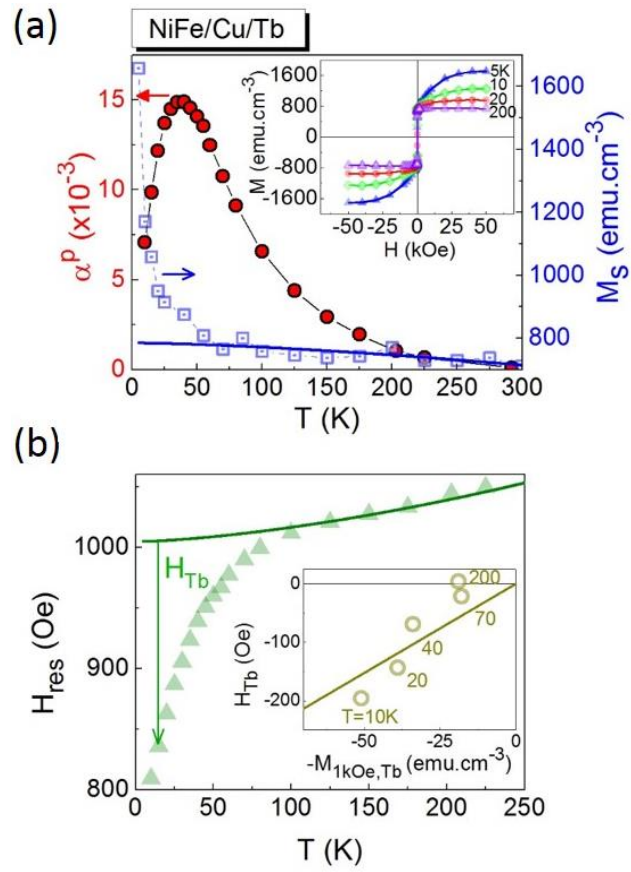

O. Gladii et al                                                                                                          Fig. 2

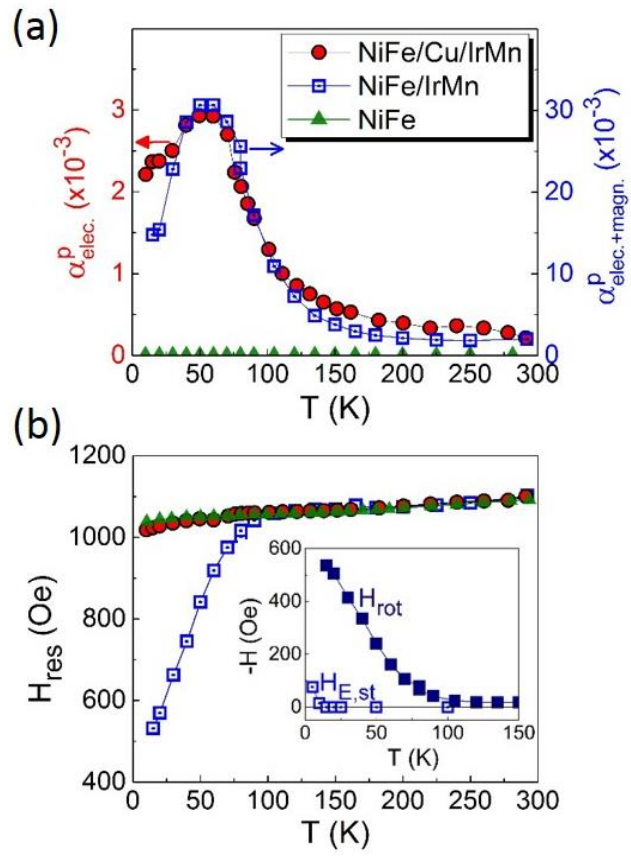

O. Gladii et al    Fig. 3

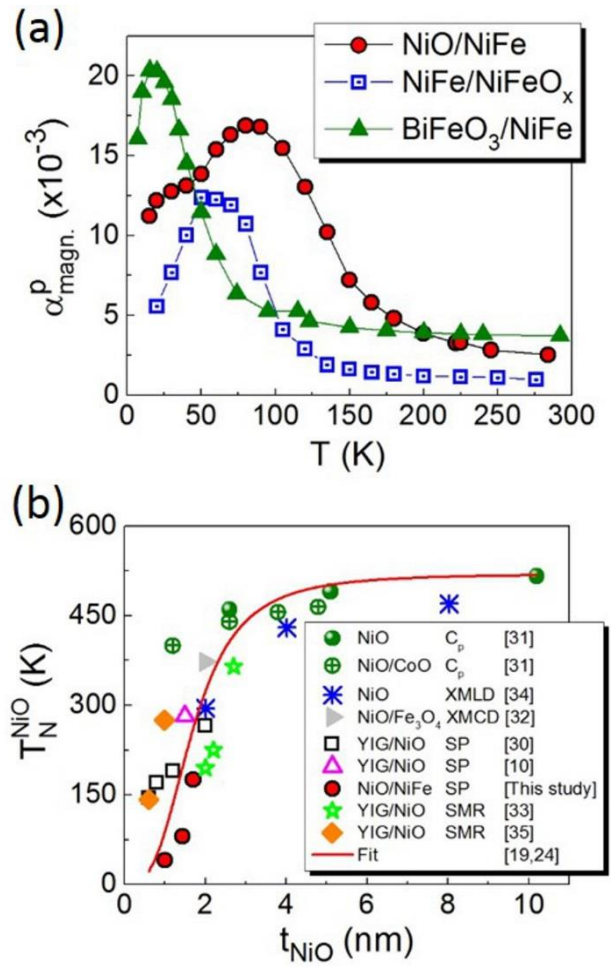

O. Gladii et al                                                                                           Fig. 4

| Stack (nm) | Nature of the spin current probe | Nature of the ordering transition | $\alpha^{ref}_{300K}$ x $10^{-3}$ | $\alpha^{p}_{300K}$ x $10^{-3}$ | $\alpha^{p}_{T=Tc}$ x $10^{-3}$ | $\frac{g^{\uparrow\downarrow}_{T=Tc}}{g^{\uparrow\downarrow}_{300K}}$ |
|---|---|---|---|---|---|---|
| NiFe(8)/**Cu(3)/Tb(3)** | Electronic, through Cu | Para. to ferro. | 10.1 | ~0 | 15 | / |
| NiFe(8)/**Cu(3)/IrMn(0.6)** | Electronic, through Cu | Para. to antiferro. | 8.1 | 0.2 | 2.9 | 14.5 |
| NiFe(8)/**IrMn(0.6)** | Electronic & magnonic | Para. to antiferro. | 8.1 | 2 | 31 | 15.5 |
| NiFe(8)/**NiFeOx(1.6)** | Magnonic | Para. to antiferro. | 8.1 | 1 | 12.3 | 12.3 |
| **NiO(1.5)**/NiFe(7) | Magnonic | Para. to antiferro. | 7.2 | 2.5 | 16.8 | 6.7 |
| **BiFeO$_3$(3)**/NiFe(8) | Magnonic | Para. to antiferro. | 8.1 | 3.6 | 20.4 | 5.7 |

O. Gladii et al                                                                 Table 1